\documentclass[prl,twocolumn,showpacs]{revtex4}
\usepackage{amsfonts,amsmath,mathrsfs,epsfig,amsbsy,bm,verbatim,subfigure}

\newcommand{\bra}[1]{\left\langle {#1} \right \vert}
\newcommand{\ket}[1]{\left\vert {#1} \right\rangle}

\newcommand{\ua}{\uparrow}
\newcommand{\da}{\downarrow}

\begin{document}
\title{How to realize  a robust practical Majorana chain in a quantum dot-superconductor linear array}
\author{Jay D. Sau$^1$}
\author{S. Das Sarma$^2$}
\affiliation{
$^1$ Department of Physics, Harvard University, Cambridge, Massachusetts 02138, USA
$^2$Condensed Matter Theory Center, Department of Physics, University of Maryland, College Park, Maryland 20742-4111, USA}

\begin{abstract}
Semiconducting nanowires in proximity to superconductors are  
promising experimental systems for Majorana fermions, which may ultimately be used as building blocks for topological quantum
computers. A serious challenge in the experimental realization of the Majorana fermions is the 
supression of topological superconductivity by disorder. We show that Majorana fermions protected by a robust 
topological gap can occur at the ends of a chain of quantum dots connected by $s$-wave superconductors. 
In the appropriate parameter regime, we establish that the quantum dot/superconductor system is equivalent to a 1D Kitaev chain, which  
  can be tuned to 
be in a robust topological phase with Majorana end modes even in the case where the  quantum dots and superconductors
are both strongly disordered. Such a spin-orbit coupled quantum dot - $s$-wave superconductor array  
provides an ideal experimental platform for the observation of non-Abelian Majorana modes.
\end{abstract}

\maketitle

\paragraph{Introduction:}
Solid state Majorana fermions (MFs) with non-Abelian statistics are  the most promising candidates for 
realizing systems with a topologically degenerate ground state. Such particles have attracted  
recent attention both due to their fundamental interest as a new type of particle with non-Abelian statistics 
and their potential application in topological quantum computation
 \cite{nayak_RevModPhys'08,Wilczek-3,levi,science}.
Over the past few years it has been realized that topological superconductors 
provide one of the conceptually simplest platforms for the solid state realization of  MFs.
More recently it was shown that  topological superconductors with MFs 
can occur fairly generically in systems containing the three ingredients of conventional $s$-wave superconductivity, 
time-reversal breaking (\textit{e.g.} via a magnetic field), and spin-orbit coupling  \cite{sau}.
A simple topological superconducting (TS) system supporting MFs, which has attracted serious 
experimental attention \cite{levi}, consists of a
 semiconductor nanowire in a  
magnetic field placed on an ordinary superconductor
 \cite{roman}.
 The $s$-wave proximity effect on a
 InAs quantum wire, which also has a sizable SO coupling,
 has already been realized in experiments \cite{doh}, and the search for the corresponding TS phase, in the presence of
 a magnetic field,  with localized Majorana modes in semiconductors is actively being pursued in many laboratories worldwide.
The theoretical demonstration ~\cite{alicea} that such 1D systems can show non-Abelian statistics has made these systems 
viable components for quantum information processing.

While the proposal for the realization of MFs in semiconductor structures has generated considerable  theoretical interest and 
even inspired significant experimental efforts \cite{expts}, a few key challenges relating to non-idealities in realistic systems
 remain.
 One such practical challenge is whether it is possible to control the chemical potential of a semiconductor 
wire, which is absolutely essential to tune the system into the TS phase, since it is in contact with a superconductor.
The other key challenge arises from the fact that the TS gap is typically suppressed by strong disorder \cite{potter}. 
Therefore the present proposals for realizing MFs would either require clean semiconductor systems or  
a spin-orbit coupling which is large enough to dominate over the disorder scattering effects \cite{potter}.
It is completely unclear at this stage whether these key obstacles of chemical potential tuning and low disorder can be overcome in the
 currently studied semiconductor nanowire structures unless substantial  materials development occurs first.  Our proposal in this paper
 overcomes both of these problems completely in one stroke by coming up with a new architecture (i.e. topology) for
 creating the Majorana-- additionally, our proposal involves using semiconductor quantum dots coupled to superconducting grains which
 is in many ways a simpler system to work with than the earlier two-dimensional \cite{sau} or one-dimensional structures \cite{roman}
 for observing the TS phase.  We are fairly confident that the  system proposed in the current paper is by far the most ideal
 experimental platform to search for the solid state Majorana modes.

 \begin{figure}[tbp]
\includegraphics[width=1.0\columnwidth]{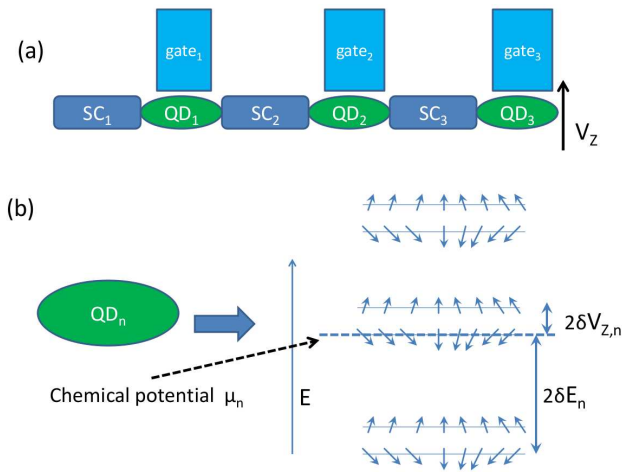}
\caption{(a) Geometry to create disorder robust TS systems with end MFs using a QD-SC array. In 
the appropriate parameter regime non-Abelian MFs are predicted to occur and be localized at the end QDs 
of a disordered QD/SC array. Each QD/SC island in the array has a different index to indicate that each QD or SC is  
 can have a different local disorder potential, hopping 
amplitude or  proximity-induced superconducting pairing potential. The chemical potential $\mu_n$ in each QD is controlled
 by the gate$_n$. Tunneling through the superconductor is assumed to be weaker than the magnetic field induced Zeeman splitting
 $\delta V_{Z,n}\sim V_Z$ of the QD levels. Applying 
fluxes to the SC islands can compensate relative signs between anomalous and normal tunneling. 
(b) Level structure of the $n^{th}$ QD  relative to the fermi level (shown by a dashed line). The level spacing $\delta E_n$ 
is assumed to be the largest 
energy scale ($\sim 2-3$ meV), the spin splitting $\delta V_{Z,n}$ is of the order $\sim 0.3$ meV. The spin texture in each level,
 which is a result 
of spin-orbit coupling and is determined by the length of the QD relative to the spin-orbit length $l_{SO}$ 
(see Eq.~\ref{eq:spintexture}), plays a crucial role in allowing spin-singlet proximity-effect between 
different QD$_n$ and QD$_{n+1}$. This level structure is qualitatively independent of local disorder potential. }
\label{Fig1}
\end{figure}

 In this paper we describe a new system  composed of a linear quantum-dot(QD)/superconductor (SC) array 
 (shown in Fig.~\ref{Fig1}(a)) to realize non-Abelian Majorana fermion modes at the ends of the array. 
The effects of disorder in the QD/SC array are suppressed in a transparent way because the topological state
 can be formed as long as the electron can hop from one QD to a neighboring QD in the array, which is a much 
easier condition to satisfy than the near ballistic transport condition necessary through a wire.
Strong disorder can typically renormalize the hopping across QD or SC islands, however as long as the disorder energy scale 
is less than the fermi energy, one can choose the QD/SC to be shorter than the localization length to enable significant 
transmission probability. The confinement potential  of the QD separates
 the  electronic energy levels in the dots as shown in Fig.~\ref{Fig1}(b),
 so that there are two levels (because of time-reversal symmetry induced Kramers-degeneracy). 
For clean TS systems, it is known that gapped superconductors are topological as long as they 
have an odd number of occupied bands at the fermi-level \cite{roman, kitaev}.  To realize a TS QD/SC array, 
the Kramer-degeneracy of the levels must be broken so that an odd number of occupied levels per QD occurs at the fermi level.   
The Kramer's degeneracy can be lifted as shown in Fig.~\ref{Fig1}(b) 
by breaking time-reversal symmetry by an externally applied magnetic field-induced Zeeman splitting $V_Z$. 
Disorder in the QD affects the wave-function and causes fluctuations in the spacing of the levels in the dots, 
but does not affect the qualitative structure shown in Fig.~\ref{Fig1}(b). In the limit where  the 
QD/SC tunneling $t$ (which is physically characterized by the inverse tunneling time from QD to SC) 
is small compared to the Zeeman splitting between the spin-split levels, the chemical potential 
can be adjusted by using gate voltages (shown in Fig.~\ref{Fig1}(a)) so that only a single level in each QD participates in   
 transport through the QD/SC array structure. As in the TS phase of other \cite{sau,roman} superconductor-semiconductor structures, 
spin-orbit coupling plays a crucial role in determining the superconducting gap in the TS phase. Because the 
single levels in each QD are spin-polarized almost in the same direction, spin-orbit coupling is necessary to lead to a 
rotation of the spin-polarization between the left and right end of the QD (as shown in Fig.~\ref{Fig1}(b)) so as to 
allow a proximity-induced SC coupling between neighboring QDs. Note that the on-site proximity-induced SC coupling in the same 
QD is forbidden by the Pauli exclusion principle. Furthermore, Coulomb charging effects can be ignored in the QD since 
we are operating in the parameter regime containing a single active level per QD.
We mention that the number of electrons per dot (they could be as few as a few 10s), whether the dots are identical or not, whether
 hopping through the whole linear array is coherent or not, how many dots there are in the system (as few as 5-10 should work)
 are all non-issues for the realization of the TS phase (as we have explicitly verified numerically)-- the only constraints are
 that there should be only one electron at the chemical potential in each dot (which can be secured by adjusting the local gates
 individually on each dot) and there should be enough hopping to ensure local proximity effect.

While for perfectly periodic QD/SC arrays, the condition for realizing a TS systems is given by whether there are an odd number
of partially filled levels in each QD, one must understand the Bogoliubov-de Gennes (BdG) Hamiltonian of the QD/SC array
 more generally to understand the
 condition for the emergence of MFs in the disordered QD/SC array. In the limit of weak (compared to level spacing in the QD)
 QD/SC tunneling, the low-energy part of the BdG quasiparticle spectrum in this system, 
 containing the zero-energy Majorana modes  in the TS phase, can 
be described in terms of an effective 1D Kitaev chain \cite{kitaev} lattice-model 
\begin{equation}
H_{eff}=\sum_n -\mu_n\hat{\psi}_n^\dagger\hat{\psi}_n+t_n(\hat{\psi}_{n+1}^\dagger\hat{\psi}_n+h.c)
+\Delta_n(\hat{\psi}_{n+1}^\dagger\hat{\psi}_n^\dagger+h.c),\label{eq:Kit}
\end{equation} 
where $\hat{\psi}_n$ is the creation operator in the active level in each QD, $\mu_n$ the effective chemical potential in each QD,
 $t_n$ and $\Delta_n$ are the normal and anomalous tunneling amplitudes between the QDs. Note that in a disordered QD/SC array,
 the amplitudes $t_n,\Delta_n$ would 
vary along the length of the chain. The complex conjugation symmetry of the full BdG Hamiltonian for the 
QD/SC system will be shown to typically result in real effective parameters $t_n,\Delta_n$.

The effective Kitaev chain description of the QD/SC array allows us to reduce the complexity of an array of disordered QDs and SCs to 
just three parameters $\mu_n,t_n,\Delta_n$ for each QD. The energy of the QD level  $\mu_n$ relative to the Fermi level 
can be controlled by a gate voltage so that only a single level in each QD is partially occupied. In the perfectly periodic case 
where the amplitudes $t_n$ and $\Delta_n$ and the local potential $\mu_n$ are independent of $n$, the topological 
condition is given by $|\mu_n|<|t_n|$\cite{kitaev}. In a disordered QD/SC array, the magnitudes of $t_n$ can be controlled
to a certain degree by controlling the QD/SC tunneling. However, the magnitude of $\Delta_n$ or the 
signs of $t_n,\,\Delta_n$ are more difficult to control given the disorder in the SC islands and the short fermi wave-vectors 
of electrons in the SC islands. In principle, such a strongly 
disordered Kitaev chain is expected to have a large number of low-energy states, similar to a disordered topological 
superconductor, hindering the observation of non-Abelian statistics and consequently topological quantum computation.
 We will show that these low-energy states can be eliminated by shifting the phase of the superconducting 
order parameters on a sub-set of the SC islands by $\pi$. By applying the appropriate phase shifts a robust TS phase with 
end MFs and a bulk gap of order $\sim \Delta_n$ with MFs at the ends can be obtained in the disordered 1D QD/SC array whenever the 
 conditions $|\mu_n|\sim 0$  and $\textrm{sign}(t_n\Delta_n)=\textrm{sign}(t_{n+1}\Delta_{n+1})$ are satisfied. More generally, 
we show in the appendix that localized MFs are obtained whenever the localization length of an effective Hamiltonian is longer than 
the superconducting coherence length.  Since $\Delta_n$ and $t_n$ are of the same order this turns out to be easy to satisfy as 
demonstrated by the exact diagonalization of large number of short Kitaev chains.
Phase shifts applied to the SC islands by values other than $\pi$ allow us to transform the Kitaev chain to this form even for complex 
hopping $t_n$ and pairing $\Delta_n$. 
Any segment of the linear QD/SC array can be tuned to the non-topological phase by tuning the chemical potential in 
the segment so that  $|\mu_{n+1}|>\textrm{max}(|t_n|,|\Delta_n|)$.
 Therefore we will establish that a gate-tunable robust TS  phase 
can be obtained in a disordered linear QD/SC array as long as disorder does not suppress any of the tunneling parameters $t_n$ to being 
vanishingly small. The TS phase would have a robust gap that would be determined by $\Delta_n$ with non-Abelian MFs 
at the ends of the QD/SC array, thus making this linear QD/SC array an ideal practical platform for the realization of non-Abelian
 statistics and MFs.

\paragraph{Spin-orbit coupled quantum dots}
To derive the effective Kitaev chain model given in Eq.~\ref{eq:Kit} for the QD/SC array one must first understand the interplay of 
confinement, Zeeman splitting, spin-orbit coupling, and disorder in determining the wave-function of the
 active level in the QD in the vicinity of the fermi-level.
Such an understanding is obtained by considering a Rashba spin-orbit coupled 1D system with a  Hamiltonian
\begin{equation}
[-\frac{1}{m^*}\partial_x^2+i\alpha \sigma_y \partial_x + +V_Z\sigma_z+V(x)]\psi(x)=E\psi(x)
\end{equation}
 where $m^*\sim 0.04 m_e$ is the effective mass in InAs, $V_Z$ is the magnetic-field induced Zeeman potential,
 $V(x)$ represents the combination of the confinement potential and  local disorder potential and $\alpha$ is the strength of 
the Rashba spin-orbit coupling. We have approximated the QD to be one-dimensional since the confiment energy of InAs QDs along the 
transverse direction can be as large as $\sim 40$ meV \cite{loss}. The Rashba spin-orbit coupling leads to a variation of the
 local spin-density along the axis $x$ of the QD, which can be described by transforming each eigenstate according to a
 spin-dependent gauge transformation  
\begin{equation}
\psi(x)\rightarrow \tilde{\psi}(x)=e^{i \pi\sigma_y x/l_{SO}}\psi(x),\label{eq:spintexture} 
\end{equation}
where $l_{SO}=\frac{\pi}{m^*\alpha}$ is the effective spin-orbit length, which is the length travelled by an electron before 
it precesses by $\pi$ under the influence of the spin-orbit magnetic field.
Substituting this into the Schrodinger equation  we find that the spin-orbit term can be completely cancelled in this new basis  
 so that the new Schrodinger equation is written as 
\begin{align}
&[-\frac{1}{m^*}\partial_x^2 +V_Z\{\sigma_z \cos{\frac{\pi x}{l_{SO}}}+\sigma_x \sin{\frac{\pi x}{l_{SO}}}\}+V(x)]\tilde{\psi}(x)\nonumber\\
&=(E+\frac{1}{m^* l_{SO}^2})\tilde{\psi}(x)\label{eq:SEpsit}.
\end{align}
Ignoring the relatively small Zeeman splitting $V_Z\sim 0.5$ meV, the energy splitting for levels in the QD $\delta E\sim \frac{\pi v_F}{2 a}$,
where $v_F\sim \sqrt{2  E/m^*}$ is the mean fermi veloctiy at an energy $E$ relative to the bottom of the band and $2 a$ is the length  of 
the QD. Since the energy $E$ must be less than the inter sub-band separation $\sim 40$ meV, we choose $E\sim 5$ meV, so that 
$\delta E\sim 3$ meV for a QD of length $2 a\sim 120$ nm. In the absence of a Zeeman splitting (i.e. $V_Z=0$), the energy levels in the QD,
 which are separated by around $\delta E\sim 3$ meV on average, are two-fold degenerate due to time-reversal symmetry.
 Disorder potentials can lead to statistical variations in the spacings of 
the QD levels. However, by tuning the gate voltage of the QD it should be possible to tune to a pair of levels that are separated 
from other levels by the average separation of $\delta E\sim 3$ meV. While the disorder potential in the QD is allowed to be large, we will 
assume that the length of the QD, $2 a$, is not significantly longer than the mean-free path so that the spatial part of the wave-function $\psi_0(x)$ 
of the pair of levels in the QD extends over the entire length of the QD. 

 Since  $V_Z\sim 0.5$ meV $\ll \delta E = 3$ meV, the effect of
 $V_Z$ can be treated as a perturbation splitting the two-fold degeneracy of the pair of states at the fermi level. 
For $a< l_{SO}$, the Zeeman field perturbation proportional to $V_Z$ in Eq.~\ref{eq:SEpsit} in the degenerate space is given by 
an effective magnetic field $V_Z[\sigma_z\bra{\psi_0}\cos{\frac{\pi x}{l_{SO}}}\ket{\psi_0}+\sigma_x\bra{\psi_0}\sin{\frac{\pi x}{l_{SO}}}\ket{\psi_0}]$ 
  for the pair of states at the fermi level. 
The Zeeman field splits the degenerate pair of levels into non-degenerate levels separated by $\delta V_Z=V_Z\sqrt{|\bra{\psi_0}\cos{\frac{\pi x}{l_{SO}}}\ket{\psi_0}|^2+|\bra{\psi_0}\sin{\frac{\pi x}{l_{SO}}}\ket{\psi_0}|^2}$, which we assume to be larger than $\delta V_Z> \frac{V_Z}{2}\sim 0.25$ meV
 for $a<l_{SO}$. Moreover the wave-function of the lower of the spin-split states is $\tilde{\psi}(x)=\psi_0(x)e^{i\eta\sigma_y}\ket{\sigma_z=-1}$ where 
$\tan{\eta}=\bra{\psi_0}\cos{\frac{\pi x}{l_{SO}}}\ket{\psi_0}/\bra{\psi_0}\sin{\frac{\pi x}{l_{SO}}}\ket{\psi_0}$. Substituting back the wave-function 
$\tilde{\psi}(x)$ into Eq.~\ref{eq:spintexture}, we obtain the approximate expression for the wave-function of the active state in the QD 
\begin{align}
&\psi_\sigma(x)=\psi_0(x)\bra{\sigma_z=\sigma}e^{i(\frac{\pi x}{l_{SO}}+\eta)\sigma_y}\ket{\sigma_z=-1}\label{eq:QDwfn}
\end{align}
where $\sigma=\pm 1$ are eigenstates of $\sigma_z$.

\paragraph{QD/$s$-wave superconductor proximity effect}
The realization of MFs in the QD system requires the introduction of the superconducting proximity effect from neighboring 
$s$-wave SC islands. Let us now consider how the proximity effect can by induced from the
 SC islands in the QD/SC chain in Fig.~\ref{Fig1}(a). Since the states in the SC are gapped we can integrate out  
the states in the SC and replace the SC by an effective Hamiltonian \cite{robustness}, which is schematically written as 
\begin{equation}
\Sigma_{SC}(E\sim 0)=t G_{SC}(E\sim 0) t^\dagger\label{eq:sigmasc}
\end{equation}
where $E\ll \Delta$ ($\Delta$ being the superconducting gap in the SC island)
 is the BdG quasiparticle energies we are solving for, $G_{SC}$ is the effective superconductor Green
 function and $t$ is the tunneling between the QDs and the SCs.
The Green-function of the SC (for uniform $\Delta$), even in the presence of disorder, is written as 
\begin{equation}
G_{SC}(xx';E=0)=\sum_n \frac{\phi_n(x)\phi_n(x')}{\epsilon_n^2+\Delta^2}\left(\begin{array}{cc}\epsilon_n&\Delta\\\Delta&-\epsilon_n\end{array}\right)\label{eq:gsc},
\end{equation} 
where $\phi_n(x)$ are the spatial parts of the electronic eigenstates with energy $\epsilon_n$ relative to the fermi-level 
in the normal state of the SC island and 
we have used the choice $(u_{\ua}(x),u_{\da}(x),v_{\da}(x),-v_{\ua}(x))^T$ for the Nambu spinor 
so that the BdG Hamiltonian for the $s$-wave superconductor is spin-independent. Here $u_\sigma(x)$ and $v_{\sigma}(x)$ 
are the electron and hole part of the BdG quasiparticle wave-functions.  The qualitative form for $G_{SC}$ is valid even for 
strongly disordered $s$-wave superconductors and the only role of strong disorder in the SC islands is to reduce the 
coherence length of the superconductor, which limits the width of the SC since the SC island is required to be shorter than the 
coherence length so that the matrix elements of $\Sigma_{SC}\propto G_{SC}(xx')$ is non-vanishing between neighboring 
QDs. In particular, for states at energies much smaller than the quasiparticle gap $\Delta$ in the superconductors, one can assume $E\approx 0$,
 so that $G_{SC}(E\sim 0)$ is particle-hole symmetric, Hermitean and spin-independent.
The effective self-energy induced by the superconductor in coordinate (i.e. $x,x'$) space can be written as a spin-independent 
sum of a normal and anomalous part as   
\begin{equation}
\Sigma_{SC}(xx')=\Sigma_{SC}^{(A)}(xx')\tau_x+\Sigma_{SC}^{(N)}(xx')\tau_z,
\end{equation}
where $\Sigma_{SC}^{(N,A)}(xx')$ are the normal and anomalous components of the self-energy.
Here the matrices $\tau_{x,y,z}$ are Pauli matrices in the Nambu space containing the particle-hole degrees of freedom.
To calculate the effective BdG Hamiltonian in the sub-space of the low-energy (i.e. near the fermi-level at $E\sim 0$)
active levels in each QD, we note that the low-energy basis of the $n^{th}$ QD consists of an electron-like wave-function 
described by an $x$-dependent 4 component spinor wave-function $\Psi_{e,n}(x)=(\psi_{n}(x),0)^T$ and a hole wave-function $\Psi_{h,n}(x)=(0,i\sigma_y \psi_n^*(x))^T=(0,i\sigma_y\psi_n(x))^T$ 
(since $\psi_n(x)$ is real), where $\psi_n(x)$ is the position dependent 2-component spinor wave-function of the active level in the QD given by 
Eq.~\ref{eq:QDwfn}.
Note that the $i\sigma_y$ matrix flips the spin of the electron spinor relative to the holes. 

Taking the relevant matrix elements of the tunneling operator $t$ of the $n^{th}$ SC island between 
the right end of the $n^{th}$ QD and the left end of the $(n+1)^{th}$ QD we find that the effective normal 
tunneling between QD$_n$ and QD$_{n+1}$ is given by 
\begin{align}
&t_n=\bra{\Psi_{e,n}}\Sigma_{SC,n}\ket{\Psi_{e,n+1}}\sim \bra{\phi_{0,n}}\Sigma_{SC,n}^{(N)}\ket{\phi_{0,n+1}}\nonumber\\
&\cos{(\frac{\pi (a_n+a_{n+1})}{l_{SO}}+\eta_n-\eta_{n+1})}
\end{align}
where we have used the QD wave-functions from Eq.~\ref{eq:QDwfn} and $\bra{\phi_{0,n}}\Sigma_{SC,n}^{(N)}\ket{\phi_{0,n+1}}$ is 
the spatial part of the matrix element of the QD wave-function with the superconducting self-energy $\Sigma_{SC,n}^{(N)}$.
Similarly, the anomalous hopping or cross-Andreev reflection amplitude induced between the dots is given by
\begin{align}
&\Delta_n=\bra{\Psi_{e,n}}\Sigma_{SC,n}\ket{\Psi_{h,n+1}}\sim \Sigma_{SC,n}^{(A)}(a_n,-a_{n+1})\nonumber\\
&\sin{(\frac{\pi (a_n+a_{n+1})}{l_{SO}}+\eta_n-\eta_{n+1})},
\end{align} 
where $\bra{\phi_{0,n}}\Sigma_{SC,n}^{(A)}\ket{\phi_{0,n+1}}$ is 
the spatial part of the matrix element of the QD  wave-function with the superconducting self-energy $\Sigma_{SC,n}^{(A)}$.
In the limit of weak barrier transparency, when the tunneling matrix elements are tuned 
to be of order $\lesssim \Delta$, we expect the matrix elements $\bra{\phi_{0,n}}\Sigma_{SC,n}^{(N,A)}\ket{\phi_{0,n+1}}$ to be 
of order of the tunneling matrix elements \cite{robustness}, which can be made of order $\Delta$.
Note that combining the above equation with Eq.~\ref{eq:QDwfn}, leads to the fact that in the limit of vanishing Rashba  spin-orbit coupling $\alpha$,
$l_{SO}$ diverges, leading to a vanishing $\eta_n$ and therefore $\Delta_n$. Thus the anomalous hopping amplitude $\Delta_n$ is proportional 
to the spin-orbit coupling. An analogous calculation leads to the conclusion that the on-site contribution (i.e. from QD$_n$ to QD$_n$) 
of the anomalous self-energy vanishes, while the on-site contribution of the normal self-energy just renormalizes $\mu_n$ by an 
amount that can be off-set by changing the gate voltage on gate$_n$ in Fig.~\ref{Fig1}(a).

\paragraph{Disordered Kitaev chain}
The resulting effective-Hamiltonian for the QD/SC array with a single active level in  $\psi_n(x)$ in QD$_n$ can be represented
 conveniently 
as a second-quantized BCS Hamiltonian of the form of Eq.~\ref{eq:Kit}. In the effective Hamiltonian $H_{eff}$, the sign of the hopping 
parameters $t_n,\Delta_n$  associated with SC$_n$, can be flipped by a gauge transformation $\hat{\psi}_m^\dagger\rightarrow -\hat{\psi}^\dagger_m$ for 
all $m >n$. Similarly, $t_n$ and $\Delta_n$ associated with SC$_n$ can be interchanged by the particle-hole transformation
 $\hat{\psi}_m^\dagger\rightarrow \hat{\psi}_m$ for all $m>n$, which is accompanied by an additional
 $\mu_m\rightarrow -\mu_m$, $t_m\rightarrow -t_m$ and $\Delta_m\rightarrow -\Delta_m$. Moreover, even in the case that certain types of 
spin-flip scattering processes (which have not being considered here in the previous paragraphs) lead to complex $t_n$, the appropriate 
phase rotations $\hat{\psi}_m\rightarrow \hat{\psi}_m e^{i\theta_n}$ can be used to make $t_n>0$.
 Therefore, without loss of generality, we can choose a gauge where $|t_n|>|\Delta_n|$ together with $t_n>0$.

As mentioned earlier, in the perfectly periodic clean limit $H_{eff}$ is in the topological phase for $|\mu_n|<|t_n|$. In 
this limit, a sign-flip of $\Delta_m$ for all $m>n$ creates a $\pi$-junction, which constitutes a domain-wall with a 
pair of localized zero-energy MF modes \cite{kitaev}. A configuration where $\Delta_n$ has the opposite sign for a block of superconducting islands 
SC$_n$ constitutes a pair of domain-walls, which confine a pair of low-energy states whose energy goes to zero exponentially in 
the length of the blocks of flipped signs in $\Delta_n$. Of course, having a gapped superconducting phase does not 
guarantee a TS phase. In fact, in general determining whether a disordered one-dimensional superconductor 
is gapped requires a calculation of the transmission matrix between the ends of the one-dimensional wire \cite{anton}. 
In what follows we will consider both analytically and numerically, whether  disordered Kitaev chains which are free 
of sign disorder in $\Delta_n$ and are otherwise random in terms of $\Delta_n$, $t_n$ and $\mu_n$ support a TS  phase with zero-energy MFs localized at
 the ends. We will refer to such Kitaev chains as being sign-ordered.

 \begin{figure}[tbp]
\includegraphics[width=.9\columnwidth]{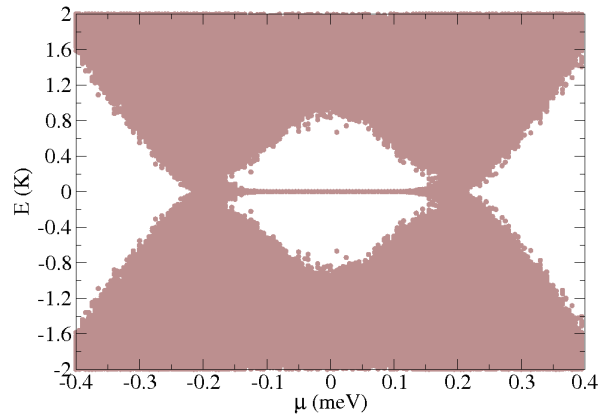}
\caption{Spectrum of a disordered 1D Kitaev chain of length $N=20$ sites.  The hopping amplitudes $t$ and pairing amplitudes $\Delta$
 are distributed randomly and uniformly  in the interval $[0.5,1.5]$ K. The spectrum shows a near-zero-mode separated by a gap 
$\sim 1$ K in a range of on-site QD chemical potential $|\mu|<0.15 meV$.
Thus the TS state shows a robust gap in this regime despite the large fluctuations in the hopping and pairing 
potential. }
\label{Fig2}
\end{figure}
\begin{figure}[tbp]
\includegraphics[width=.9\columnwidth]{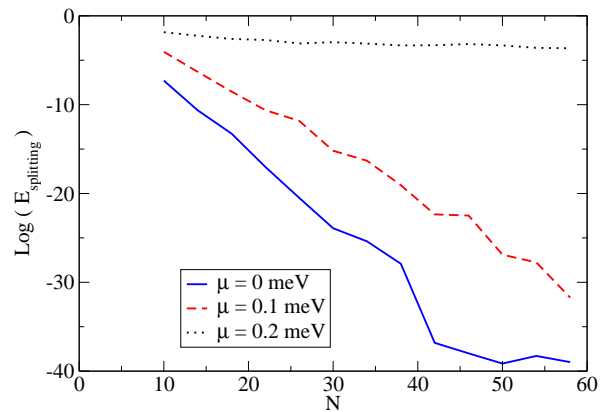}
\caption{Finite size splitting of Majorana modes as a function of chain-length. Finite chain length leads to overlap of 
end MFs which splits them away from zero energy in any finite system. The MF splitting $E_{splitting}$
 in the disordered 1D Kitaev chain (with the same parameters as in Fig. 2) is found to go to zero exponentially in chain length $N$. }
\label{Fig3}
\end{figure}

In the supplementary material, using a transfer matrix formalism similar to Ref.~\cite{anton}, we show that the Kitaev chain 
with real parameters considered here has a chiral symmetry and therefore can be mapped to the problem of the existence of 
zero modes at the ends of a non-Hermitean Hamiltonian with an imaginary vector potential ~\cite{directed_localization}. 
In particular, we show that a long sign-ordered Kitaev chain has MF zero modes at the end with a gap in the bulk whenever
 $\textrm{sgn}(t_n\Delta_n)=\textrm{sgn}(t_{n+1}\Delta_{n+1})$ (i.e. sign ordered)
  and the chemical potential in each QD has been tuned 
using a gate voltage so that $\mu_n=0$. In the case of $|\mu_n|>0$, end modes can be shown to exist whenever 
\begin{equation}
\zeta=E\textrm{log}\{\frac{|t_n-\Delta_n|}{t_n+\Delta_n}\}+\zeta_0<0 \label{eq:topcond}
\end{equation}
where the expectation value is taken with respect to the random distribution for the magnitudes of $|t_n|,|\Delta_n|$ and $\zeta_0^{-1}$
 is the localization length of an effective hermitean Hamiltonian with on-site potential $\mu_n$ and hopping $\sqrt{t_n^2-\Delta_n^2}$. 
 The MF zero mode is localized at the ends with a localization length of $|\zeta|^{-1}$ sites. The parameter regime $\zeta<0$ supporting 
end MFs corresponds to the delocalized phase of an effective non-Hermitean Hamiltonian $H_0+\Delta$ (defined in the supplementary material)
~\cite{directed_localization}.
 The other parameter regime with $\zeta>0$ corresponds to the localized phase ~\cite{directed_localization} of the non-hermitean Hamiltonian
 $H_0+\Delta$  so that 
both the non-hermitean Hamiltonian $\tilde{H}_0$ and the Kitaev chain $H_{eff}$ have low-energy modes in the bulk. 
  Furthermore, in the supplementary material we provide a detailed discussion of why the sign-ordered Kitaev chain with
 the Hamiltonian in Eq.~\ref{eq:Kit} remains gapped for otherwise arbitrary disorder under this condition.

While the analytic results provide some understanding of how zero-energy MF arise in the sign-ordered Kitaev chain, it is 
more insightful to consider the spectra of typical realization of finite sign-ordered Kitaev chains. We consider  
in Fig.~\ref{Fig2} the spectrum of a 20-site (we have used several different number of sites in our numerical work
 getting the same qualitative answer)
 Kitaev chain with open boundary conditions. Choosing a fixed value for the gate voltage $\mu_n$ 
across the chain  and  realistic random values of $t_n$ and $\Delta_n$, which vary over a significantly larger range then their minimum values,
 we find a robust TS phase with end MFs, which appear as near zero-energy eigenstates.
 The end MFs are not strictly localized at the ends of the chain but are split by a finite amount due to overlap of the decaying MF wave-functions 
across the chain. The exponential dependence on the chain length of the energy splitting seen in Fig.~\ref{Fig3} shows clearly the exponential 
localization of the MFs and also demonstrates that the protection of the topological degeneracy of the MFs increases exponentially 
with the length of the chain.

The relative signs of the hopping and anomalous tunneling amplitudes $t_n$ and $\Delta_n$ arise the randomness on the scale of the fermi wave-vector 
$k_F$ of the electronic wave-functions $\phi_n(x)$ in the SC islands. Of course, one cannot in general expect to eliminate the random relative sign 
arising from such short scale randomness. This problem can be remedied by adjusting the  individual phases of each superconducting island SC$_n$,
which according to Eqs.~\ref{eq:sigmasc} and ~\ref{eq:gsc} provides an independent handle on the signs of $t_n$ and $\Delta_n$.

\paragraph{Conclusion}
In conclusion, the two major problems of the semiconductor superconductor structures, proposed \cite{sau,roman} 
in the context of creating non-Abelian
 solid state Majorana modes, are disorder and gating. Both problems are solved by 
using the relatively simple QD/SC linear array structure shown in Figure~\ref{Fig1}.
 The use of SC grains can potentially introduce random $\pi$ junctions 
in the chain, which can produce sub-gap states.  These effects of strong disorder are avoided by adding an array of 
gates and removing accidental $\pi$ junctions by phase shifting the superconductors. 
In addition, the chemical potential in each dot is tuned individually with gates, thus completely eliminating any gating problem
 of the semiconductor.  Finally, structures similar to the QD/SC array being proposed in this work are already in use in various
 conventional quantum computing architectures such as quantum dot spin qubits and superconducting Cooper pair box qubits,
 thus making this a relatively practical (as well as extremely robust) proposal for creating non-Abelian modes in the laboratory.

This work is supported by Microsoft Q, DARPA-QuEST and the Harvard Quantum Optics Center.

\appendix
\section{Supplementary material}
\subsection{Zero energy MFs at the end of a 1D Kitaev chain}
Consider a 1D Kitaev chain Hamiltonian 
\begin{equation}
H_{eff}=\sum_n -\mu_n\hat{\psi}_n^\dagger\hat{\psi}_n+t_n(\hat{\psi}_{n+1}^\dagger\hat{\psi}_n+h.c)+\Delta_n(\hat{\psi}_{n+1}^\dagger\hat{\psi}_n^\dagger+h.c).
\end{equation} 
In this sub-section we prove that if $t_n,\Delta_n$ are real and sign-ordered i.e.
\begin{equation}
\textrm{sign}(\Delta_n t_n)=\textrm{sign}(\Delta_{n+1} t_{n+1})\label{eq:condition},
\end{equation} 
and $|\mu_n|<\textrm{max}(|t_{n-1}|,|\Delta_{n-1}|)$, then the Kitaev chain has zero-energy MFs at the ends.
As discussed in the main text, we can choose $|t_n|>|\Delta_n|$ and $t_n>0$ without loss of generality. In the topological regime 
where Eq.~\ref{eq:condition} is satisfied, $\Delta_n$ has the same sign on all sites in the chain.

Defining a Nambu spinor $(\psi_n^\dagger,\psi_n)$, with a corresponding particle-hole symmetry operator $\Lambda=\tau_x K$, the 
BdG Hamiltonian corresponding to the above BCS Hamiltonian is written as 
\begin{equation}
H_{BdG}=H_0\tau_z+i\Delta\tau_y,
\end{equation} 
where 
\begin{align}
&H_{0}=\sum_n -\mu_n\ket{n}\bra{n}+t_n[\ket{n}\bra{n+1}+h.c]\\
&\Delta=\sum_n \Delta_n[\ket{n}\bra{n+1}-h.c].
\end{align}
Applying a unitary transformation $U=\frac{1-i\tau_y}{\sqrt{2}}$, $H_{BdG}$ transforms into
\begin{equation}
UH_{BdG}U^\dagger=\left(\begin{array}{cc}0&(H_0+\Delta)\\(H_0+\Delta)^T&0\end{array}\right).
\end{equation} 
The BdG equation 
\begin{equation}
\left(\begin{array}{cc}0&(H_0+\Delta)\\(H_0-\Delta)&0\end{array}\right)\left(\begin{array}{c}\zeta_1\\\zeta_2\end{array}\right)=E\left(\begin{array}{c}\zeta_1\\\zeta_2\end{array}\right)
\end{equation} 
is now written as 
\begin{align}
&(H_0+\Delta)\zeta_2=E\zeta_1\\
&(H_0-\Delta)\zeta_1=E\zeta_2.\label{eq:chiralBdG}
\end{align}
The matrices $H_0\pm\Delta$ are written as 
\begin{align}
&H_{0}\pm\Delta=\sum_n -\mu_n\ket{n}\bra{n}\nonumber\\
&+\sum_n (t_n\pm\Delta_n)\ket{n}\bra{n+1}+(t_{n-1}\mp\Delta_{n-1})\ket{n}\bra{n-1}.
\end{align}
The above Hamiltonians are non-Hermitean and display localization transitions similar to the Hamiltonians 
in previous work ~\cite{directed_localization}.
Zero energy (i.e. $E=0$) modes are obtained as solutions of one of the two decoupled equations
\begin{equation}
(H_0\pm\Delta)\zeta_{2,1}=0.
\end{equation}
The sign of $\Delta_n$ can be inverted by flipping the chain, therefore in this basis if one of the ends has a 
MF mode with $\zeta_2=0$ then the other end has an MF mode with $\zeta_1=0$.
Since both signs of $\Delta$ occur in the above equation, we can assume that $\Delta_n>0$ without loss of generality. 

Writing the wave-function as $\psi=\sum_n \psi_n\ket{n}$, the equation for the zero-mode can be written as a transfer-matrix-like  
relation 
\begin{equation}
-\mu_n\psi_{n}+(t_n+\Delta_n)\psi_{n+1}+(t_{n-1}-\Delta_{n-1})\psi_{n-1}=0.\label{eq:psin}
\end{equation}
Redefining 
\begin{align}
&\psi_n\rightarrow \tilde{\psi}_n=g_n\psi_n\nonumber\\
&g_n=\prod_{1\leq m <n}\sqrt{\frac{(t_{m}+\Delta_{m})}{(t_{m}-\Delta_{m})}}\label{eq:tildepsi}
\end{align}
 and dividing  Eq.~\ref{eq:psin} by $g_n$, we obtain the equation as 
or equivalently as 
\begin{equation}
-\mu_n\tilde{\psi}_{n}+\sqrt{(t_n+\Delta_n)}\tilde{\psi}_{n+1}+\sqrt{(t_{n-1}^2-\Delta_{n-1}^2)}\tilde{\psi}_{n-1}=0.\label{eq:psin1},
\end{equation}
which in turn can be re-written as 
\begin{align}
&\tilde{\psi}_{n+1}=\frac{1}{\sqrt{(t_n^2-\Delta_n)}}[\mu_n\tilde{\psi}_{n}-\sqrt{(t_{n-1}^2-\Delta_{n-1}^2)}\tilde{\psi}_{n-1}]\label{eq:psin2}.
\end{align}
The above equation represents a zero-mode for a hermiteanized tight-binding Hamiltonian 
\begin{align}
&\tilde{H}_{0}=\sum_n -\mu_n\ket{n}\bra{n}\nonumber\\
&+\sum_n \sqrt{(t_n^2-\Delta_n^2)}\ket{n}\bra{n+1}+\sqrt{(t_{n-1}^2-\Delta_{n-1}^2)}\ket{n}\bra{n-1}\label{eq:tildeh0}.
\end{align}

Let us start by considering the special case of the above Hamiltonian Eq.~\ref{eq:tildeh0}, where each gate in the QD has been tuned so that $\mu_n=0$, 
so that $\tilde{H}_0$ has a chiral symmetry which supports zero-end modes similar to Su-schrieffer-Heeger model~\cite{anton1}. The  equation for 
$\tilde{\psi}_n$ then simplifies to   
\begin{align}
&\tilde{\psi}_{n+1}=-\frac{\sqrt{(t_{n-1}^2-\Delta_{n-1}^2)}}{\sqrt{(t_n^2-\Delta_n^2)}}\tilde{\psi}_{n-1},
\end{align}
together with $\tilde{\psi}_1=0$.
The above equations can be solved for $\tilde{\psi}_p$ as 
\begin{align}
&\tilde{\psi}_{2n}=(-1)^n\prod_{p<n}\frac{\sqrt{(t_{2 p}^2-\Delta_{2 p}^2)}}{\sqrt{(t_{2 p-1}^2-\Delta_{2 p-1}^2)}}\tilde{\psi}_{0}\nonumber\\
&\tilde{\psi}_{2n-1}=0.
\end{align}
In the case where the hopping amplitudes are statistically independent of each other, the pre-factor 
\begin{equation}
\prod_{p<n}\frac{\sqrt{(t_{2 p}^2-\Delta_{2 p}^2)}}{\sqrt{(t_{2 p-1}^2-\Delta_{2 p-1}^2)}}\sim e^{\pm \sqrt{n}\sigma^2}
\end{equation}
in the large $n$ limit where 
\begin{equation}
\sigma^2=\textrm{var}(\textrm{log}(\sqrt{(t_{p}^2-\Delta_{p}^2)})),
\end{equation}
where the variance $\textrm{var}$ is taken with respect to the random distribution.
Since the pre-factor for $\tilde{\psi}_{2 n}$ scales as $e^{\sqrt{n}}$, and the factor $g_n$ in the definition of 
$\tilde{\psi}_n$ scales as $g_n\sim e^{-n\zeta_1}$ where 
\begin{equation}
\zeta_1=-E\textrm{log}\left\{\frac{|t_n-\Delta_n|}{t_n+\Delta_n}\right\},\label{eq:zeta1}
\end{equation}
and $E$ is the expectation value with respect to the distribution of $t_n,\Delta_n$,
it follows that 
\begin{equation}
\psi_{2 n}\sim e^{\pm\sqrt{n}\sigma^2-2 n\zeta_1}\sim e^{-2 n\zeta_1},
\end{equation}
so that the zero-mode wave-function $\psi_{2 n}$ is localized near the end of the chain and is normalizable.
This shows that, as claimed in the main text that localized zero-energy MF modes exist for $\mu_n=0$, for independently 
ordered bonds. However, one should note that such localized MFs don't exist for every configuration of $t_n,\Delta_n$.
For example, a dimerized configuration with $t_{2n -1}>t_{2 n}$ would not 
lead to zero modes below a critical value of $\Delta_n$.

Now we discuss the more general case for $\mu_n\neq 0$. In this case the solution $\tilde{\psi}_n$ of the transfer matrix relation Eq.~\ref{eq:psin2} 
is expected to scale as 
\begin{equation}
\tilde{\psi}_n\sim e^{n\zeta_0}
\end{equation}
where $\zeta_0^{-1}$ is the localization length for Eq.~\ref{eq:tildeh0}. Using the definition of $\psi_n$ in terms of $\tilde{\psi}_n$ we find that 
\begin{equation}
\psi_n\sim  e^{n\zeta}
\end{equation}
where $\zeta=(\zeta_1-\zeta_0)$, proving Eq.~\ref{eq:topcond} so that $\psi_n$ is normalized and localized when $\zeta<0$.
 This regime corresponds to the delocalized phase 
of Ref.~\cite{directed_localization}. Note that this condition is also necessary, since in the localized phase one will generically 
obtain low-energy localized states in $\tilde{H}_0$, which will lead to similar energy states in $H_0+\Delta$~\cite{directed_localization}.
This establishes the conditions where one can rigorously expect localized end modes with a gap. However, the numerics show that even for short 
chains and $\mu_n\neq 0$, localized MFs exist.

\subsection{Lower bound on the gaps of sign-ordered Kitaev chains}
While we have shown the existence of Majorana fermions at the ends of the Kitaev chain for a large set of parameters, 
the thermal robustness of the topological phase is determined by the quasiparticle gap in the system with periodic boundary 
conditions. To estimate the smallest possible (worst case) gap of $H_{BCS}$, we need to find a lower bound on $E^2$ to the solutions 
of Eq.~\ref{eq:chiralBdG} for sign-ordered Kitaev chains in the delocalized phase when Eq.~\ref{eq:topcond} is satisfied.

It follows from Eq.~\ref{eq:chiralBdG} that $E^2$ are eigenvalues of both the real-symmetric matrices 
\begin{align}
&(H_0+\Delta)^T(H_0+\Delta)\zeta_2=E^2\zeta_2\\
&(H_0+\Delta)(H_0+\Delta)^T\zeta_1=E^2\zeta_1.
\end{align}
Since it is easier to prove bounds upper bounds on the maximum eigenvalues then lower bounds on the minimum eigenvalue,
 we consider the inverse matrices 
  \begin{align}
&(H_0+\Delta)^{-1,T}(H_0+\Delta)^{-1}\zeta_1=E^{-2}\zeta_1\\
&(H_0+\Delta)^{-1}(H_0+\Delta)^{-1,T}\zeta_2=E^{-2}\zeta_2.
\end{align}
It is well-known from linear-algebra~\cite{linear_algebra}, that the maximum value of $E^{-2}$ is related
 to the maximum row sum  of the matrix $(H_0+\Delta)^{-1}$,i.e.
\begin{align}
&E^{-2}=\frac{\zeta_1^T (H_0+\Delta)^{-1,T}(H_0+\Delta)^{-1}\zeta_1}{\zeta_1^T\zeta_1}\nonumber\\
&<\textrm{max}_j\sum_i |((H_0+\Delta)^{-1} )_{i,j}|\textrm{max}_i\sum_j |((H_0+\Delta)^{-1} )_{i,j}|.
\end{align} 
Therefore we get the lower bound on the magnitude of the eigenvalues 
\begin{equation}
|E|> \sqrt{\left(\textrm{max}_j\sum_i |((H_0+\Delta)^{-1} )_{i,j}|\textrm{max}_i\sum_j |((H_0+\Delta)^{-1} )_{i,j}|\right)}.\label{eq:Ebound}
\end{equation}
Matrices whose inverses decay exponentially and are finite have energies that are bounded away from zero.

The inverse of $(H_0+\Delta)$ can be computed using the transfer matrix form for Eq.~\ref{eq:psin2} 
 can be written as a matrix recursion relation 
\begin{equation}
\Psi_n= \left(\begin{array}{c}0\\\frac{\tilde{\phi}_n}{\sqrt{t_n^2-\Delta_n^2}}\end{array}\right)+\left(\begin{array}{cc}0&1\\-\frac{\sqrt{(t_{n-1}^2-\Delta_{n-1}^2)}}{\sqrt{t_n^2-\Delta_n^2}}&-\frac{\mu_n}{\sqrt{t_n^2-\Delta_n^2}}\end{array}\right)\Psi_{n-1},
\end{equation}
where $\Psi_n=(\tilde{\psi}_{n},\tilde{\psi}_{n+1})^T$.
To calculate the inverse of $H_0+\Delta$ for a chain with periodic boundary conditions, we can label any site to be $0$ and choose
 $\phi_0=1$ and $\phi_{n\neq 0}=0$. Furthermore, we will make the ansatz $\Psi_{-1}=0$. This leads to the sequence of equations 
\begin{align}
&\Psi_0=\left(\begin{array}{c}0\\\frac{1}{\sqrt{t_0^2-\Delta_0^2}}\end{array}\right)\nonumber\\
&\Psi_{n>0}=\left(\begin{array}{cc}0&1\\-\frac{\sqrt{(t_{n-1}^2-\Delta_{n-1}^2)}}{\sqrt{t_n^2-\Delta_n^2}}&-\frac{\mu_n}{\sqrt{t_n^2-\Delta_n^2}}\end{array}\right)\Psi_{n-1}.
\end{align}
Using arguments from the theory of 1D localization using transfer matrices $\Psi_{n>0}$ is expected to grow as $\Psi_n\sim e^{n\zeta_0}$, where $\zeta_0^{-1}$ is the localization length of the hermitean Hamiltonian in Eq.~\ref{eq:tildeh0}. By transforming back to the original $\psi_n$ variables
 using Eq.~\ref{eq:tildepsi}, we find using Eq.~\ref{eq:zeta1}  that in the delocalized phase 
\begin{equation}
\psi_n\sim  \frac{1}{\sqrt{t_0^2-\Delta_0^2}}e^{n\zeta}
\end{equation}
with $zeta<0$ so that for long Kitaev chains the corrections related to our Ansatz $\Psi_{-1}=0$ are exponentially small.
Furthermore this proves a bound of $\sim \frac{1}{\Delta_0 }$ on the row-sums of $\textrm{max}_i\sum_j |((H_0+\Delta)^{-1} )_{i,j}|$ in Eq.~\ref{eq:Ebound}. On the other hand, by considering the transpose $(H_0+\Delta)^T$, one can obtain a similar bound on the column sum.
Substituting in Eq.~\ref{eq:Ebound}, we obtain a gap of order $\sim|\Delta_n|$ for the Kitaev chain. Of course this is a rough estimate since we 
have not carefully estimated values for the localization length $\zeta_0^{-1}$. However, the numerical results in Fig.~\ref{Fig2} and ~\ref{Fig3} 
seem to be consistent with these expectations.

\end{document}